\documentclass[]{IEEEtran}
\usepackage{blindtext}
\usepackage{graphicx}
\usepackage{xcolor}
\usepackage{hyperref}

\usepackage[]{textgreek}

\usepackage[alsoload=synchem]{siunitx}
\DeclareSIUnit\pixel{pixel}
\DeclareSIUnit\electron{e^{-}}
\DeclareSIUnit\adu{ADU}
\DeclareSIUnit\count{c}

\ifCLASSINFOpdf
\else
\fi
\usepackage{tikz}

\newcommand\copyrighttext{%
  \footnotesize \textcopyright 2017 IEEE. Personal use of this material is permitted.  Permission from IEEE must be obtained for all other uses, in any current or future media, including reprinting/republishing this material for advertising or promotional  purposes, creating new collective works, for resale or redistribution to servers or lists, or reuse of any copyrighted component of this work in other works.
  DOI: \href{https://dx.doi.org/10.1109/TNS.2017.2762281}{10.1109/TNS.2017.2762281}}
\newcommand\copyrightnotice{%
\begin{tikzpicture}[remember picture,overlay]
\node[anchor=south] at (current page.south) {\fbox{\parbox{\dimexpr\textwidth-\fboxsep-\fboxrule\relax}{\copyrighttext}}};
\end{tikzpicture}%
}

\usepackage[cmex10]{amsmath}
\DeclareMathOperator\e{e}

\begin{document}
\title{Optimal Pulse Processing, Pile-up Decomposition and Applications of Silicon Drift Detectors at LCLS}

\author{G.~Blaj*,~\IEEEmembership{Member,~IEEE},
        C.~J.~Kenney,~\IEEEmembership{Member,~IEEE},
        A.~Dragone,
        G.~Carini,~\IEEEmembership{Member,~IEEE},
        S.~Herrmann,
        P.~Hart,
        A.~Tomada,~\IEEEmembership{Member,~IEEE},
        J.~Koglin,
        G.~Haller,~\IEEEmembership{Member,~IEEE},
        S.~Boutet,
        M.~Messerschmidt,
        G.~Williams,
        M.~Chollet,
        G.~Dakovski,
        S.~Nelson,
        J.~Pines,
        S.~Song,
        J.~Thayer% <-this % stops a space
\thanks{Manuscript received January~26,~2017; revised June~3,~2017 and September~9,~2017; accepted October~4, 2017. SLAC-PUB-16991.}%
\thanks{G.~Blaj, C.~J.~Kenney, A.~Dragone, G.~Carini, S.~Herrmann,  P.~Hart, A.~Tomada, J.~Koglin, G.~Haller, S.~Boutet, M.~Messerschmidt, G.~Williams, M.~Chollet, G.~Dakovski, S.~Nelson, J.~Pines,  S.~Song, and J.~Thayer  are with SLAC National Accelerator Laboratory, Menlo~Park,~CA~94025, U.S.A.}%
\thanks{A.~Tomada is currently with the SRI International, Menlo Park,~CA~94025, U.S.A., M.~Messerschmidt is currently with the European~XFEL, 22869~Schenefeld, Germany and G.~Williams is currently with Brookhaven National Laboratory, Upton, NY 11973, U.S.A.}% <-this % stops a space
\thanks{* Corresponding author: blaj@slac.stanford.edu.}%
}


\markboth{}%
{Shell \MakeLowercase{\textit{et al.}}: Bare Demo of IEEEtran.cls for Journals}



\maketitle
\copyrightnotice

\begin{abstract}
%\boldmath
Silicon drift detectors (SDDs) revolutionized spectroscopy in fields as diverse as geology and dentistry. For a subset of experiments at ultra-fast, x-ray free-electron lasers (FELs), SDDs can make substantial contributions. Often the unknown spectrum is interesting, carrying science data, or the background measurement is useful to identify unexpected signals. Many measurements involve only several discrete photon energies known a priori, allowing single event decomposition of pile-up and spectroscopic photon counting. We designed a pulse function and demonstrated that the signal amplitude (i.e., proportional to the detected energy and obtained from fitting with the pulse function), rise time, and pulse height are interrelated and at short peaking times the pulse height and pulse area are not optimal estimators for detected energy; instead, the signal amplitude and rise time are obtained for each pulse by fitting, thus removing the need for pulse shaping. By avoiding pulse shaping, rise times of tens of nanoseconds resulted in reduced pulse pile-up and allowed decomposition of remaining pulse pile-up at photon separation times down to hundreds of nanoseconds while yielding time-of-arrival information with precision of 10~nanoseconds. Waveform fitting yields simultaneously high energy resolution and high counting rates (2 orders of magnitude higher than current digital pulse processors).
At pulsed sources or high photon rates, photon pile-up still occurs. We showed that pile-up spectrum fitting is relatively simple and preferable to pile-up spectrum deconvolution. We developed a photon pile-up statistical model for constant intensity sources, extended it to variable intensity sources (typical for FELs) and used it to fit a complex pile-up spectrum. We subsequently developed a Bayesian pile-up decomposition method that allows decomposing pile-up of single events with up to 6 photons from 6 monochromatic lines with 99\% accuracy. The usefulness of SDDs will continue into the x-ray FEL era of science. Their successors, the ePixS hybrid pixel detectors, already offer hundreds of pixels, each with similar performance to an SDD, in a compact, robust and affordable package.

\end{abstract}

\begin{IEEEkeywords}
Silicon drift detectors, pulse processing, free-electron lasers, pulse pile-up, photon pile-up, Bayesian decomposition, x-ray spectroscopy, photon counting
\end{IEEEkeywords}

%\ifCLASSOPTIONpeerreview
%    \begin{center} \bfseries EDICS Category: 3-BBND \end{center}
%\fi
\IEEEpeerreviewmaketitle

\section{Introduction}
%Text width: \the\textwidth, column width: \the\columnwidth.
\subsection{Silicon Drift Detectors at Free-Electron Laser Sources}
Silicon drift detectors (SDDs) \cite{gatti1984semiconductor} are a well-established technology that has revolutionized spectroscopy in fields as diverse as geology and dentistry. At a first glance it would seem that detectors with such a slow response would not be suitable for the new ultra-fast x-ray free-electron lasers (FEL) coming online \cite{emma2010first}. However, FELs require a range of detectors with different specifications to cover the applications space \cite{graafsma2009requirements,blaj2014detector,blaj2015xray}. For a subset of experiments at FELs, SDDs can make substantial contributions \cite{nasri2015front, blaj2016performance}.

Free-electron laser (FEL) pulses are typically very short (in the order of tens of femtoseconds). LCLS typically produces pulses at a rate of \SI{120}{\hertz}. While each FEL pulse can contain \numrange{E12}{E13} x-ray photons, the experiment design (e.g., optics, sample, detector position, filters,  collimators) can reduce the number of detected photons to several per pulse.

Photon pile-up with usual sources (e.g., x-ray tubes, scanning electron microscopes) typically occurs at different time intervals, resulting in changes of the waveform shape. However, at FEL sources, the signal from photons detected in a small detector area (e.g., through a small collimator aperture onto an SDD) arrive at nearly the same time. This results in a waveform with the same shape as when detecting individual photons, and a pulse height corresponding to their aggregated energy.

While photons detected at different radii in an SDD arrive at nearly the same time, different radii result in different drift times and consequently, in changes in the waveform shape.

Many measurements involve only several distinct photon energies known a priori, significantly simplifying photon pile-up analysis. Often the unknown spectrum is interesting, carrying science data, or the background measurement is useful to identify unexpected signals.

We investigate the performance of silicon drift detectors at x-ray FELs. In particular we study the ability to decompose the pile-up spectrum that results from various combinations of a few wavelengths and the possibility of separately recording photons that are absorbed at different radii (thus having varying drift times).

\subsection{Pulse Processing}
The typical approach to pulse processing in detection with a wide range of detectors (scintillation, gas, high purity germanium, SDD, transition edge) is to preamplify the detector signal, perform pulse shaping, and collect histograms of the resulting pulse peak heights \cite{wilson1950noise}.

When individual waveforms are (partially) overlapping, pulse pile-up occurs, modifying pulse height and resulting in spectrum distortion \cite{wielopolski1976prediction}. This can be classified in two broad classes: (1)~long term, ``tail'' pile-up, resulting in resolution degradation (further called ``pulse pile-up''); and (2)~coincidences (overlapping in the peaking part), resulting mainly in sum distortion, further called ``photon pile-up''. 

Both pile-up mechanisms occur with increasing probability for larger pulse shaping and decay times. Short peaking times result in a noise increase (e.g., for SDDs, increasing hyperbolically with decreasing peaking times \cite{redus2009digital}).

To mitigate this problem, often pulse processors have a slow channel for energy resolution and a fast channel for time resolution and pile-up rejection \cite{rozen1961pile}. Signals arriving within short time intervals will not be separated by the pile-up rejection; instead, they result in photon pile-up. Considerable effort has been invested on mathematical descriptions of spectrum distortion in the presence of pile-up and pile-up rejection (e.g., \cite{wielopolski1976prediction, gardner1977generalized1}) and spectrum correction (e.g., \cite{wielopolski1977generalized2, nakhostin2010digital}).

Current commercial pulse processors typically perform pulse shaping with digital filters in the frequency domain \cite{gatti1990optimum}, reducing costs and improving stability compared to analog pulse shaping, while taking advantage of current field-programmable gate arrays (FPGAs).

As an alternative to pulse shaping of scintillator signals, \cite{guo2005study} predicted dispensing with analog processing and using fast ADC conversion followed by software waveform fitting.

By eliminating pulse shaping, the rise time can be much shorter, significantly reducing pile-up. Pile-up pulses can be fitted in real time with a fixed, expected pulse shape \cite{scoullar2011real}; pulse shape variation results in significant residuals and rejection \cite{scoullar2011real}. This approach yields energy and time-of-arrival of individual photons in multi-pulse pile-up while enabling higher rates of detection (e.g., $\approx \SI{190}{\kilo\count\per\second}$ with $\SI{10}{\percent}$ rejection) \cite{scoullar2011real}.

For SDDs, different drift times result in different amounts of diffusion, and consequently, different rise times. In the absence of pulse shaping, these differences in rise times are significant. We show that the signal amplitude (i.e., proportional to the detected energy and obtained from fitting with the pulse function), rise time and pulse height are interrelated; variation in rise times of individual pulses must be taken into account in a function fit for optimal energy resolution in the absence of pulse shaping. With this approach, pulse rejection is not necessary, and we achieve a pulse separation of $\approx\SI{730}{\nano\second}$ (corresponding to $\approx\SI{1.4}{\mega\count\per\second}$).

Computing power, memory, storage and data transfer rates continue to improve, making it likely that pulse processing will continue to transition towards full software processing (taking advantage of increasing computation resources and fulfilling the promise of largely avoiding pulse pile-up).

\subsection{Pile-up Reduction}
We define ``one photon peaks'' as the peaks in the energy histogram (spectrum) corresponding to detection of single photons, ``two photon peaks'' as peaks corresponding to the aggregated energy resulting from detection of the pile-up of two photons (with the same energy or not), etc. We also define the ``average photon rate'' $\lambda$ as the number of photons in all events divided by the number of events, and will often refer to it as ``photon rate''.

As a consequence of the Poissonian counting statistic, the one photon peak area is a non-injective, nonlinear, peak-shaped function of the corresponding average photon rate. For low rates, the peak area can be used to approximate the photon rate directly. At higher photon rates, the dependency becomes nonlinear, and beyond a certain point (average rate $\lambda=1$~photon/event) decreases. Measuring higher average photon rates accurately benefits from using the pile-up peaks instead of deconvolving them.

We present here an accurate model for photon pile-up of discrete spectra with increasing complexity, from monochromatic through bicolor to general and use it for fitting a discrete spectrum, effectively deconvolving it. We further expand the model to include variable intensity sources with a gamma distribution, typical for, e.g., FEL sources.

Other approaches to single event Bayesian decomposition of photon pile-up use complex iterative Markov chain Monte Carlo \cite{van2004highly} or $\chi^2$ fitting \cite{davis2001event}. Both are optimized for x-ray astronomy, expecting limited pile-up, complex spectra, and iteratively analyzing each event, updating the spectrum, and repeating until convergence. They are highly complex and relatively slow.

Using the photon pile-up model we developed a Bayesian pile-up decomposition method which yields the individual photon energies for each photon pile-up event and evaluate its accuracy to \num{99}{\percent} for a spectrum with photon pile-up of up to \num{6} photons from \num{6} monochromatic lines.

\section{Methods}

\subsection{Experimental Set-up}
\begin{figure}[!t]
\centering
\includegraphics[width=\columnwidth]{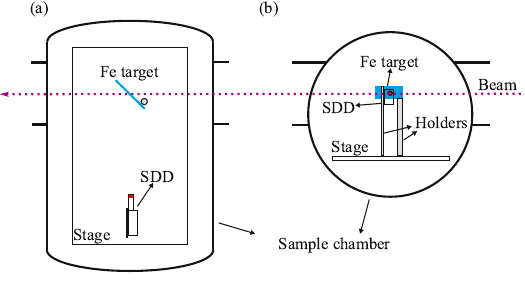}
% where an .eps filename suffix will be assumed under latex, 
% and a .pdf suffix will be assumed for pdflatex; or what has been declared
% via \DeclareGraphicsExtensions.
\caption{Schematic diagram of the sample chamber at the Coherent X-ray Imaging instrument at LCLS: (a)~top view: the direct beam passes through an Fe target and exits the chamber; fluorescence and scattered photons are collected at a right angle onto the silicon drift detector (SDD); the entire setup is enclosed in the vacuum chamber; (b)~side view, showing the Fe target and SDD at beam height.}
\label{fig1}
\end{figure}
The experiments were performed at the Linac Coherent Light Source (LCLS) at SLAC National Accelerator Laboratory. Two LCLS instruments were used, X-ray Pump-Probe (XPP) \cite{chollet2015xray} and Coherent X-ray Imaging (CXI) \cite{boutet2010coherent}. Fig.~\ref{fig1} shows a schematic diagram of the CXI experimental chamber, with a Fe target and an SDD detector.

We projected FEL beam pulses of $\approx$~\SI{9}{\kilo\electronvolt} photons on an Fe target (at FELs, the photon energy and beam intensity are somewhat variable from pulse to pulse), resulting in a detected signal composed of scattered \SI{9}{keV} radiation together with Fe fluorescence (Fe~K\textalpha~at \SI{6.40}{\kilo\electronvolt} and Fe~K\textbeta~at \SI{7.06}{\kilo\electronvolt}).

We used a standard Amptek~XR-100SDD detector with a resolution of \SI{125}{\electronvolt} full width at half maximum (FWHM) at \SI{5.9}{\kilo\electronvolt}, an active area of \SI{25}{\square\milli\metre} and thickness of \SI{500}{\micro\metre}. 

To observe and characterize the differences between the central and peripheral areas of the SDD, we mounted in front of the SDD a collimator with two cylindrical apertures (\SI{0.5}{\milli\metre} diameter), placed over the SDD such that one aperture was aligned with the center of the SDD and the other was at a \SI{2.5}{\milli\metre} radius towards the periphery of the SDD. We will subsequently refer them as the ``central'' and ``peripheral'' apertures.

Using collimators to convert standard SDDs to position sensitive detectors comes at the cost of reduced fill factor in our experiment, with each each aperture passing about \SI{0.8}{\percent} of the detector area. However, a large number of photons was present and the reduced fill factor was not an issue. In practice, apertures can be useful to limit the number of photons reaching the detector.

\subsection{Waveform Acquisition and Processing\label{sec2b}}

The SDD preamplifier output was connected to the Amptek PX5 pulse processor, with the SDD current amplified and high-pass filtered in the PX5 preamplifier and prefilter blocks, respectively; for details and conceptual diagram see \cite{amptek2016px5}. The analog prefilter output of the pulse processor was routed to the PX5 AUX1 analog port. 

The PX5 pulse processor does not provide the analog prefilter output signal directly; instead, this signal is digitized by the PX5 ADC at \SI{80}{\mega\hertz}, subsequently synthesized by the PX5 DAC at \SI{10}{\mega\hertz}, and then routed to the PX5 AUX1 analog output; see details and conceptual diagram in \cite{amptek2016px5}. In Appendix~\ref{appendix} we describe the method used for recovery of ADC input waveforms from the raw measured waveforms.

The analog signal waveforms were then acquired by an Acquiris high-speed digitizer, with \num{8192} samples at \SI{100}{\mega\hertz} for each FEL pulse and triggered pulse by pulse by the low jitter LCLS trigger signal. We will use the term ``events'' to refer to individual acquisitions from single FEL pulses. Each event resulted in one single waveform. We acquired \num{354304} events and corresponding waveforms at \num{120}~events per second.

\subsection{Pulse Function\label{sec2c}}

The pulse height analysis can be optimally performed in the frequency domain \cite{gatti1990optimum}. Here we chose to perform the analysis in the time domain, allowing accurate reconstruction of timing, clipping and pile-up of pulses.

\begin{figure}[!t]
\centering
\includegraphics[width=\columnwidth]{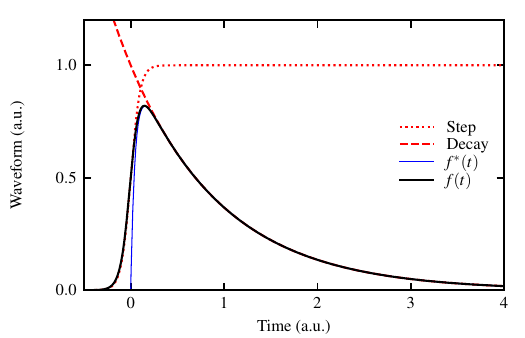}
% where an .eps filename suffix will be assumed under latex, 
% and a .pdf suffix will be assumed for pdflatex; or what has been declared
% via \DeclareGraphicsExtensions.
\caption{Example of pulse function $f(x)$ (solid black line, Eq.~\ref{eq2}) obtained by multiplying a gradual step function (dotted red line) with an exponential decay (dashed red line); for reference, a commonly used function with exponential peaking $f^*(t)$ is shown with a thin blue line.}
\label{fig2}
\end{figure}

An individual pulse has a shape characterized by a relatively rapid increase near the arrival time $t_0$ and a relatively slow exponential decay towards zero afterwards. An often used pulse function is:
\begin{align}  \label{eq1}
    f^*(t)=A \left(1-\e^{-\frac{t-t_0}{\tau_p}}\right) \e^{-\frac{t-t_0}{\tau_d}}
\end{align}
which assumes that the peaking occurs as an exponential decay towards \num{1} with peaking time $\tau_p$ (corresponding to the peaking time of pulse shaping) multiplied by an exponential decay with decay time $\tau_d$ (corresponding to the time constant of the prefilter high-pass filter). The typical pulse function $f^{*}(t)$ is illustrated in Fig.~\ref{fig2} by the thin blue line (usually the peaking time is chosen to be much longer than rise times, in the order of microseconds instead of tens of nanoseconds, leading to a much slower pulse onset).

After high-pass filtering in the analog prefilter, and in the absence of pulse shaping, peaking is fast, with a slower onset and ending resulting from charge diffusion. We can parametrize this behavior by using a smooth step function with time constant $\tau_s$ (corresponding to the SDD and preamplifier rise time):
\begin{align} \label{eq2}
   f(t)=A \frac{1}{1+\e^{-\frac{t-t_0}{\tau_{s}}}} \e^{-\frac{t-t_0}{\tau_{d}}}
\end{align}
with the step factor, decay factor and their product (i.e., pulse function $f(t)$) indicated in Fig.~\ref{fig2} by the dotted red line, dashed red line and thick black line, respectively.

\subsection{Pulse Function Characteristics\label{sec2d}}
We compare the characteristics of the typical pulse function $f^{*}(t)$ and the modified pulse function $f(t)$, relating their characteristics to the detected signal. Eq.~\ref{eq1}~and~\ref{eq2} are translation invariant; in this subsection we conveniently choose a time origin $t_0=0$ to simplify the analysis of the pulse function.

\subsubsection{Pulse Height}

We can obtain the relation between pulse height $\max f(t)$ and signal amplitude $A$ by finding $t_1$ where the first derivative of the pulse function $f(t)$ is zero:
\begin{align}
\begin{split} \label{eq3}
    {f^*}'(t)=0 \Rightarrow t_1^*=\tau_s\log{\frac{\tau_d+\tau_p}{\tau_p}}\\
    f'(t) = 0 \Rightarrow t_1=\tau_s\log{\frac{\tau_d-\tau_s}{\tau_s}}
\end{split}
\end{align}
and substituting Eq.~\ref{eq3} into Eq.~\ref{eq1},~\ref{eq2}:
\begin{align}
\begin{split} \label{eq4}
    \max f^*(t)= A \frac{\tau_d}{\tau_d+\tau_p}\left(\frac{\tau_d}{\tau_p}-1\right)^{-\frac{\tau_p}{\tau_d}}\\
    \max f(t)= A \frac{\tau_d-\tau_s}{\tau_d}\left(\frac{\tau_d}{\tau_s}-1\right)^{-\frac{\tau_s}{\tau_d}}
\end{split}
\end{align}

\subsubsection{Pulse Area}
The pulse area can be calculated:
\begin{align}
\begin{split} \label{eq5}
    \int_{-\infty}^{\infty}f^*(t) dt = A \frac{\tau_d^2}{\tau_d+\tau_p}\\
    \int_{-\infty}^{\infty}f(t) dt = A \pi \tau_s \csc\left(\frac{\pi \tau_s}{\tau_d}\right) \text{ when } \tau_s < \tau_d
\end{split}
\end{align}
where $\csc$ is the cosecant function.

\subsubsection{Discussion}

In subsection~\ref{sec3b} we show that the step time (or rise time) $\tau_s$ changes with location (central or peripheral), thus changing the ratios between fitting parameter $A$, pulse height, and pulse area. However, using function $f(t)$ for pulse processing allows recovering the position, time and pulse pile-up for each event, without pulse shaping.

For $\{\tau_s, \tau_p\} \ll \tau_d$, the equations above are simplified to:
\begin{align} \label{eq6}
\begin{split}
    \max f(t)=\max f^*(t) = A\\
    \int_{-\infty}^{\infty} f(t) dt = \int_{-\infty}^{\infty} f^*(t) dt = A \tau_d
\end{split}
\end{align}
thus for long peaking and decay times, the signal amplitude, pulse height and pulse area show a similar behavior.

\subsection{Two Pulse Pile-up Decomposition\label{sec2e}}
In this experiment we use a single channel SDD, which provides a single waveform for each event; at FELs, photons arrive to the detector plane nearly at the same time. Due to the two apertures and differences in drift time between the central and peripheral areas, the waveforms contain two partially overlapping pulses with amplitudes proportional to the energy of photons detected through each aperture.

To account for the nonzero baseline, clipping of larger signals ($V>\SI{0.3}{\volt}$), and the superposition of two pulses with two different amplitudes and arrival times, we extend Eq.~\ref{eq2} to:

\begin{align} \label{eq7}
    y(t) = \operatorname{min}\Bigg(y_0 + \frac{A^c \cdot \e^{-\frac{t-t_0^c}{\tau_d}}}{1 + \e^{-\frac{t - t_0^c}{\tau_s^c}}} + \frac{A^p \cdot \e^{-\frac{t-t_0^p}{\tau_d}}}{1 + \e^{-\frac{t - t_0^p}{\tau_s^p}}}, y_{max}\Bigg),
\end{align}
where the $^c$ and $^p$ superscripts denote parameters specific for the central and the peripheral aperture, respectively.

Each individual waveform was then fitted (least squares) with the function in Eq.~\ref{eq7}, where the signals induced by photons entering through the central and peripheral apertures have amplitudes \(A^c\) and \(A^p\), arrival times \(t_0^c\) and \(t_0^p\), with 'decay' and 'step' times denoted by \(\tau_d\), \(\tau_s^c\) and \(\tau_s^p\), offset \(y_0\) and maximum (clipping) digitized voltages \(y_{max}\). Signal amplitudes for either aperture are collected in histograms.

\begin{figure}[!t]
\centering
\includegraphics[width=\columnwidth]{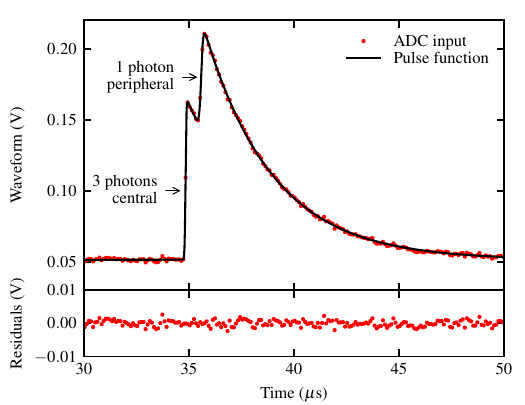}
% where an .eps filename suffix will be assumed under latex, 
% and a .pdf suffix will be assumed for pdflatex; or what has been declared
% via \DeclareGraphicsExtensions.
\caption{For each event (i.e., FEL pulse and trigger), fitting (least squares, Eq.~\ref{eq7}) central and peripheral aperture traces with their respective amplitudes (\(A^c\) and \(A^p\)), timing (\(t_0^c\) and \(t_0^p\)) information, despite pile-up or occasional clipping; this particular trace shows an example with \num{3} photons entering through the central aperture and \num{1} photon through the peripheral aperture.}
\label{fig3}
\end{figure}

Fig.~\ref{fig3} shows an example of waveform (red dots) and corresponding least squares fit (black line), with \num{3} photons entering through the central aperture and \num{1} photon entering through the peripheral aperture.

\subsection{Photon Pile-up Statistics at Constant Beam Intensity\label{sec2f}}
For a single, monochromatic line $i$ with average detection rate $\lambda_i$, the probability of a pile-up event with $j$ photons in one acquisition is given by the Poisson distribution:
\begin{align} \label{eq8}
    P_i^j=P(\lambda_i,j)=\frac{\lambda_i^j \e^{-\lambda_i}}{j!}
\end{align}
where $j$ is an exponent (unlike elsewhere in this paper, where superscript $j$ is typically an index).

\subsubsection{Peak area and detection rate}
The relationship between 1 photon peak area (i.e., detection probability $P_i^0$) and average detection rate $\lambda_i$ for 1 photon peaks is:
\begin{align} \label{eq9}
    P_i^1=\lambda_i \e^{-\lambda_i}
\end{align}
For low rates, $\lim_{\lambda_i \to 0} \lambda_i \e^{-\lambda_i} = \lambda_i$, resulting in the usual approximation:
\begin{align} \label{eq10}
    \lambda_i \approx P_i^1 \text{ for } \lambda_i \to 0
\end{align}

However, for relatively high rates in the presence of pile-up, the full form must be used. Solving Eq.~\ref{eq9} for $\lambda_i$ we obtain two solutions:
\begin{align} \label{eq11}
    \lambda_i=
    \begin{cases}
        -\W_0(-P_i^1), \lambda_i \le 1\\
        -\W_{-1}(-P_i^1), \lambda_i > 1
    \end{cases}
\end{align}
where $\W_0(x)$ and $\W_{-1}(x)$ are the principal and \num{-1} solutions provided by the Lambert-W functions.

Eq.~\ref{eq11} reflects the fact that a 1-photon peak area corresponds to two different average detection rates $\lambda_i$ on either side of $\lambda_i=1$~photon/event, thus it is a degenerate measure of photon rate. For low rates ($\lambda_i < 1$~photon/event), the 1 photon peak area can be used; however, for high rates, the pile-up peaks must be considered.

\subsection{Photon Pile-up Statistics with FEL Beams\label{sec2g}}
The self-amplified spontaneous emission process at FELs results in significant pulse-to-pulse intensity variation, often described by a gamma probability distribution function \cite{hogan1998measurements}:
\begin{align} \label{eq12}
    \Gamma_{PDF}(\alpha,\beta_\Gamma,x)=\frac{\beta_\Gamma^{\alpha}}{\Gamma(\alpha)} x^{\alpha-1} \e^{- \beta_\Gamma x}
\end{align}
where $\Gamma$ is the gamma function, with average rate $\lambda=\alpha \beta_\Gamma$ and shape parameter $\alpha$. For simplicity we replace $\beta_\Gamma$ with $\lambda/\alpha$.

The gamma distribution of FEL pulse intensities results in a ``stretching'' of the Poisson distribution corresponding to the average FEL intensity. The stretching effect resulting from the variability of the FEL beam intensity can be calculated as a weighted average of the Poisson distributions corresponding to different FEL pulse intensities, weighted by their probability:
\begin{multline} \label{eq13}
    _{\Gamma}P_i^j=\int_0^{\infty} P(x,j) \Gamma_{PDF}(\alpha,\lambda_i/\alpha,x) dx \\
    =\frac{\lambda_i^j \alpha^\alpha}{(\lambda_i+\alpha)^{\alpha+j}} \frac{\Gamma(\alpha+j)}{j! \Gamma(\alpha)}
\end{multline}
(integrated with \cite{wolfram2017mathematica}). Note that Eq.~\ref{eq13} is a generalized form of the Poisson distribution in Eq.~\ref{eq8}, which can be recovered in the limit $\lim_{\alpha \to \infty}{_{\Gamma}P_i^j} = P_i^j$. In this paper we used Eq.~\ref{eq13} where applicable; Eq.~\ref{eq8} is provided for reference and should be used for data measured with sources of constant intensity.

\subsubsection{Peak area and detection rate}
The relationship between 1 photon peak area (i.e., detection probability $P_i^0$) and detection rate $\lambda_i$ for one photon peaks results by setting $j=1$ in Eq.~\ref{eq13}:
\begin{align} \label{eq14}
    _{\Gamma}P_i^1=\lambda_i \left(\frac{\alpha}{\lambda_i+\alpha}\right)^{\alpha+1}
\end{align}
which can be solved numerically for $\lambda_i$. As in Eq.~\ref{eq11}, Eq.~\ref{eq14} yields two solutions on either side of $\lambda_i=1$~photon/event (on average), thus the photon pile-up peaks must be considered for evaluating $\lambda_i$.

\subsection{Peak Shape\label{sec2h}}

\subsubsection{Peak width dependency on energy}
At energy $E$, the noise r.m.s. is given by the electronic noise $\sigma_0$ and an energy dependent noise component (including, i.e., Fano noise) $\sqrt{\sigma_1 E}$, added quadratically:
\begin{align} \label{eq15}
    \sigma(E)=\sqrt{\sigma_0^2+\sigma_1 E}
\end{align}
Using this simplified form allows fitting peak shapes with only two global fitting parameters, $\sigma_0$ and $\sigma_1$, resulting in increased fitting stability.

\subsubsection{Normal distribution}
We used a normal distribution function for modeling peak shapes of single peaks $i$ (corresponding to the monochromatic lines at energies $E_i$):
\begin{align} \label{eq16}
    \mathcal{N}(E,E_i)=\frac{\e^{-\frac{(E-E_i)^2}{2 \sigma(E_i)^2}}}{\sqrt{2 \pi}\sigma(E_i)}
\end{align}

\subsubsection{Version 1 generalized normal distribution}
In practice we observe that the peaks display ``heavy tails''; we expect this to be specific to this experiment, see subsection~\ref{sec4b2} for details. To model heavy tails, we use a version 1 generalized normal distribution:
\begin{align}  \label{eq17}
    \mathcal{N}_1(E,E_i,\beta) = \frac{\beta \e^{-\left(\frac{|E-E_i|}{\sqrt{2}\sigma(E_i)}\right)^{\beta}}}{2 \sqrt{2} \sigma(E_i) \Gamma(\frac{1}{\beta})}
\end{align}
where $\beta$ is the shape parameter (e.g., a parameter $\beta=2$ reverts this function to the normal distribution).
 
\subsubsection{General peak shape} Combining the two distributions above we obtain the peak function:
\begin{align}  \label{eq18}
    \mathcal{N}^*(E,E_i)=(1-\eta) \mathcal{N}(E,E_i) + \eta \mathcal{N}_1(E,E_i,\beta)
\end{align}
where $\eta$ is the fraction of photons in heavy tails.

\subsection{Photon Pile-up Spectra\label{sec2i}}
In this subsection we deduce increasingly complex pile-up spectra: (1)~one photon peak spectrum, (2)~pile-up of monochromatic line, (3)~pile-up of two lines (useful in, e.g., pump-probe experiments), to (4)~pile-up of multiple lines.

Note that for all calculated spectra we use probability density functions (PDFs). To obtain histograms from PDFs we multiply the PDF value at each histogram bin position with the number of events $N$ and the histogram bin size.

\subsubsection{One photon spectrum (i.e., no pile-up)}
The 1 photon fitting function is the sum of the peaks of $L$ discrete lines and zero peak ($E_0=0$):
\begin{align} \label{eq19}
    S^1(E)=\sum_{i=0}^{L} \mathcal{N}(E,E_i) P_i^1
\end{align}
which is subsequently used to fit (least squares) the areas $A_i$ of the $N$ peaks and the background $b_j$.

\subsubsection{Pile-up of a single line (i.e., monochromatic beam)}
The general form for peaks corresponding to $j$ photons from line $i$ is:
\begin{align} \label{eq20}
    E_i^j = j E_i
\end{align}
The total number of photons detected from line $i$ can be obtained by summation over Eq.~\ref{eq16} with a resulting spectrum:
\begin{align} \label{eq21}
    S_i(E)=N \sum_{j=0}^{\infty} \mathcal{N}(E,E_i^j) P_i^j
\end{align}
where $N$ is the number of pulses.

\subsubsection{Pile-up of two lines}
The photon pile-up of two lines $\{i_1,i_2\}$ can be obtained similarly:
\begin{align} \label{eq22}
    E_{i_1 i_2}^{j_1 j_2}=j_1 E_{i_1} + j_2 E_{i_2}
\end{align}

The resulting $0$ to $M$ photon spectrum is:
\begin{align} \label{eq23}
    S_{i_1 i_2}^{0..M}=N \sum_{j_1=0}^{M} \sum_{j_2=0}^{M-j_1} \mathcal{N}(E,E_{i_1 i_2}^{j_1 j_2}) P_{i_1}^{j_1} P_{i_2}^{j_2}
\end{align}
Eq.~\ref{eq23} contains $\binom{M+2}{2}$ terms corresponding to the binomial expansion coefficient:
\begin{align} \label{eq24}
    \binom{M+2}{2}=\frac{(M+2)!}{M! 2!}
\end{align}

\subsubsection{Pile-up of multiple lines}
The photon pile-up of $L$ lines $\{i_1,i_2,..i_L\}$ is:
\begin{align} \label{eq25}
    E_{i_1 i_2 .. i_L}^{j_1 j_2 .. j_L}=\sum_{k=1}^{L} j_k E_{i_k}
\end{align}

The resulting $0$ to $M$ photon spectrum is:
\begin{align} \label{eq26}
    S_{i_1 i_2 .. i_L}^{0..M}=N \sum_{j_1=0}^{M} \sum_{j_2=0}^{M-j_1}.. \sum_{j_L=0}^{M-\sum_{k=1}^{L-1}j_{k}} \mathcal{N}(E,E_{i_1 i_2.. i_L}^{j_1 j_2.. j_L}) \prod_{k=1}^L P_{i_k}^{j_k}
\end{align}

Eq.~\ref{eq26} contains $\binom{L+M}{L}$ terms; the algorithm complexity is $O(n!)$, increasing rapidly for large numbers of lines $L$ and photons $M$.

\subsection{Pile-up Spectrum Fitting and Decomposition}

Eq.~\ref{eq26} accurately describes the photon pile-up of multiple lines and could be used for $\chi^2$ fitting of entire spectra. In spectrum fitting we added a third degree polynomial background to the theoretical spectrum:
\begin{align} \label{eq27}
    I_{bg}(E)=b_0+b_1 E+b_2 E^2 + b_3 E^3
\end{align}

The number of counts per histogram bin has a Poisson distribution with an expected error r.m.s. of $\sqrt{y}$; we add \num{1} to account for bins with zero counts, obtaining for each histogram bin an expected error $y_{err} \approx 1+\sqrt{y}$ (adimensional). This error estimate is used for weighting individual bins in spectrum fitting.

The variance of the resulting fitting parameters is the principal diagonal of the covariance matrix \cite{richter1995estimating}; its square root yields the r.m.s. error of the fitting parameter.

Eq.~\ref{eq26} allows obtaining the average photon rates $\lambda_i$ directly, eliminating the need for pile-up deconvolution; the number of photons from each line $i$ is $N \lambda_i$, where $N$ is the number of events observed. 

\subsection{Bayesian Decomposition of Single Event Photon Pile-up\label{sec2k}}
The probability that a particular energy $E$ belongs to each of the $\binom{L+M}{L}$ pile-up peaks corresponding to the combinations of $M$ photons and $L$ lines can be calculated from the individual terms of Eq.~\ref{eq26}. For example, in the the case of $L=6$, $M=6$, there are \num{924} pile-up peaks.

\subsubsection{Maximum likelihood estimation of photon pile-up decomposition\label{sec2k1}}
For each detected energy $E$, we construct a \num{924} element column vector $\mathbf{P}(E)$, with each of the \num{924} rows containing the probability that energy $E$ belongs to the \num{924} corresponding pile-up peaks, calculated from individual terms in Eq.~\ref{eq26}.

The $\mathbf{P}(E)$ row with the maximum value indicates the most likely pile-up peak (i.e., maximum likelihood estimator for the pile-up peak).

\subsubsection{Error in estimating pile-up peaks\label{sec2k2}}
The estimated error of peak assignment is
\begin{align} \label{eq28}
    Err(E)=1-\frac{\max \mathbf{P}(E)}{\sum_k \mathbf{P}_{k}(E)}
\end{align}
and the total error is the weighted average of the energy dependent $Err(E)$ with the probability density function of the energy $S(E)$:
\begin{align} \label{eq29}
    Err=\int S(E) Err(E) dE
\end{align}
This yields the error rate of pile-up peak identification.

\subsubsection{Error in estimating number of photons\label{sec2k3}}
For estimating the error rate of photon identification, we simulate \num{E+6} events, with numbers of photons sampled randomly from their respective distributions. For each simulated event we simulate an energy response by sampling one value from the corresponding distribution Eq.~\ref{eq16}, and then recontructing the individual photon energies as above. We generate a \num{6x6} confusion matrix $\mathbf{C}$, with rows representing simulated photons and columns representing assigned photons after single event pile-up reconstruction.

The error rate of single event Bayesian decomposition of photon pile-up can be calculated from the ratio between the sum of the diagonal and the sum of all cells in confusion matrix $\mathbf{C}$:
\begin{align} \label{eq30}
    Err=1-\frac{\sum_i \mathbf{C}_{i,i}}{\sum_{i_{r}} \sum_{i_{c}} \mathbf{C}_{i_{r},i_{c}}}
\end{align}

\subsubsection{Discussion}
While the error rates of the Bayesian, maximum likelihood decomposition depend strongly on the spectrum characteristics, the method presented here can be applied to estimate deomposition errors in any other spectrum.

\section{Pulse Processing Results}

\begin{figure}[!t]
\centering
\includegraphics[width=\columnwidth]{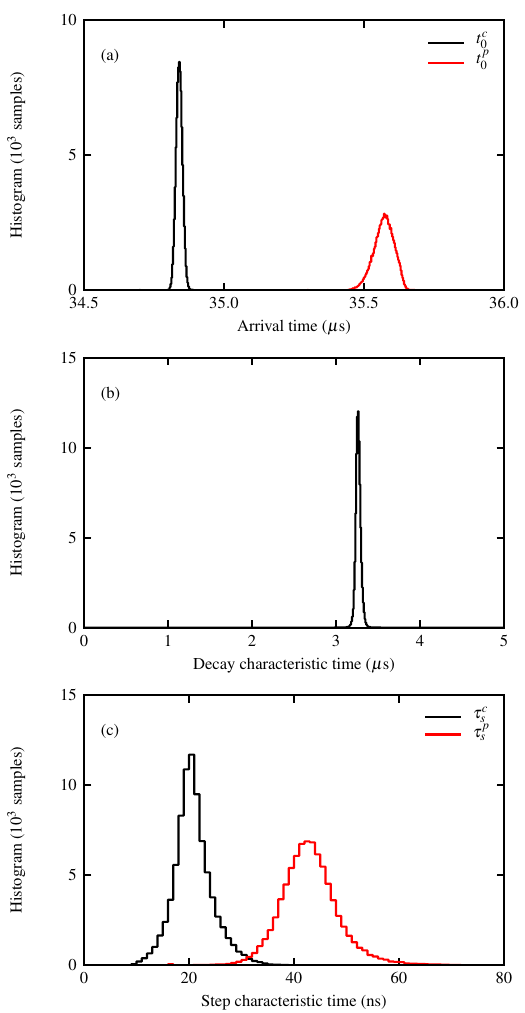}
% where an .eps filename suffix will be assumed under latex, 
% and a .pdf suffix will be assumed for pdflatex; or what has been declared
% via \DeclareGraphicsExtensions.
\caption{Distribution of fitting results: (a) shows a histogram of pulse arrival times $t_0^c=\num{34.842}\pm\SI{0.012}{\micro\second}$ (black line, central aperture) and $t_0^p=\num{35.574}\pm\SI{0.037}{\micro\second}$ (red line, peripheral aperture); at FELs, photons reach the detector surface at nearly the same time, thus the difference of \SI{732}{\nano\second} is due to the difference of charge drift times and effectively turns the SDD into a position sensitive detector with tens of independent channels and increasing resolution towards the SDD center; (b) shows the histogram of the decay characteristic time $\tau_d=\num{3.269}\pm\SI{0.035}{\micro\second}$, corresponding to the nominal \SI{3.2}{\micro\second} time constant of the high-pass filter of the analog prefilter, and (c) shows the histogram of step characteristic times $\tau_s^c=\num{21.0}\pm\SI{3.8}{\nano\second}$ (black line, central aperture) and  $\tau_s^p=\num{43.3}\pm\SI{5.9}{\nano\second}$ (red line, peripheral aperture).}
\label{fig4}
\end{figure}

Time results from fitting Eq.~\ref{eq7} are summarized in Fig.~\ref{fig4} with (a)~histograms of arrival times (central and peripheral), (b)~decay characteristic time (common) and (c)~step characteristic times (central and peripheral).

\subsection{Pulse Arrival Times and Interaction Radii}

Arrival time results are shown in Fig.~\ref{fig4}~(a); the pulse arrival times for the central aperture (black line, $t_0=\num{34.842}\pm\SI{0.012}{\micro\second}$) and peripheral (red line, $t_p=\num{35.574}\pm\SI{0.037}{\micro\second}$) are clearly separated.

As the photons reach the SDD surface nearly at the same time, the difference of \SI{732}{\nano\second} in pulse arrival times is due to the charge drift time between the SDD periphery and center \cite{segal1996simulation}. The arrival time jitter (\SI{12}{\nano\second} r.m.s. in the center and \SI{37}{\nano\second} r.m.s. in the periphery) is significantly lower than the difference of arrival times, allowing precise pulse by pulse measurement of arrival times (and corresponding interaction radii), effectively turning the SDD into a position sensitive detector with tens of independent position channels and increasing resolution towards the center.

\subsection{Decay and Step Characteristic Times\label{sec3b}}

The decay time shown in Fig.~\ref{fig4}~(b), $\tau_d=\num{3.269}\pm\num{0.035}\si{\micro\second}$, corresponds to the nominal \SI{3.2}{\micro\second} time constant of the high-pass filter of the analog prefilter \cite{amptek2016px5}, and has a relatively small variation ($\sigma$/mean~$\approx$~\SI{1.1}{\percent}).

The step characteristic times depicted in Fig.~\ref{fig4}~(c) are clearly different, with a faster step transition in the central compared to the peripheral area; as both signals are processed identically, the difference is due to the supplementary charge diffusion between the periphery and center.

\subsection{Optimal Estimator for Photon Energy\label{sec3c}}
Several estimators for detected photon energy can be considered: the pulse height, pulse area, and fitted signal amplitudes $A$ (Eq.~\ref{eq7}). In the absence of pulse shaping, step characteristic times can differ significantly (subsection~\ref{sec3b}), resulting in different behaviors of these estimators (as described in subsection~\ref{sec2d}).

For each event, the central and peripheral signal amplitudes $A^c$ and $A^p$ were obtained by fitting individual event waveforms with Eq.~\ref{eq7}. The pulse height and area were estimated from the fit parameters with Eq.~\ref{eq4} and Eq.~\ref{eq5}, respectively; using fit parameters and Eq.~\ref{eq4} to estimate pulse height significantly reduced the high frequency noise.

\begin{figure}[!t]
\centering
\includegraphics[width=\columnwidth]{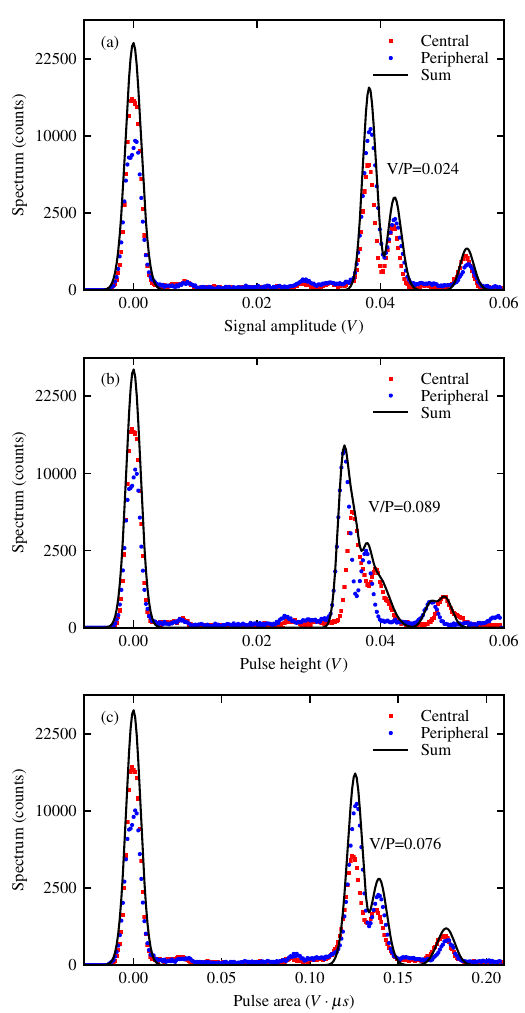}
% where an .eps filename suffix will be assumed under latex, 
% and a .pdf suffix will be assumed for pdflatex; or what has been declared
% via \DeclareGraphicsExtensions.
\caption{Histograms of (a)~fitted signal amplitude $A$ (Eq.~\ref{eq7}), (b)~pulse height estimated by peak fitting to minimize high frequency noise, and (c)~pulse area, with histograms of central pulses indicated by red squares, peripheral pulses with blue dots, and their sum with the black line; the extent of separation of the two partly overlapping peaks (Fe~K\textalpha{ }and Fe~K\textbeta) is quantified by the valley-to-peak ratio (V/P) \cite{christophe1971valley} of the sum histogram; note that with typical pulse shaping (raise times in the order of microseconds to minimize loss of energy resolution), the individual contributions of photons entering through the central and peripheral apertures can't be separated and the counting rate is limited \cite{amptek2017sddres}; the optimal estimator for photon energy in the absence of pulse shaping is the fitted signal amplitude $A$, with a V/P ratio of \num{0.024}.}
\label{fig5}
\end{figure}

In Fig.~\ref{fig5}, we show the histograms of (a)~fitted signal amplitudes, (b)~pulse height, and (c)~pulse area, with histograms of central pulses indicated by red squares, peripheral pulses with blue dots, and their sum with the black line.

The optimal energy resolution of the SDD is achieved at the best separation of the partially overlapping peaks (Fe~K\textalpha{ }and Fe~K\textbeta) of the sum histogram. The extent of separation can be quantified using the valley-to-peak ratio (V/P) \cite{christophe1971valley}, defined as the ratio of the heights of the trough between the peaks and of the highest peak, respectively. The ratios are indicated in the figure.

The optimal estimator for photon energy in the absence of pulse shaping is the fitted signal amplitude $A$, with a V/P ratio of \num{0.024}, while using the pulse area or pulse height results in higher V/P ratios (\num{0.076} and \num{0.089}, respectively) and decreased energy resolution and gain matching between the center and the periphery of the SDD.

In typical pulse processing, the dependence on peaking time is mitigated by increasing peaking time $\tau_p$ through pulse shaping (typically with a rise time in the order of microseconds), reducing the effect of variations in $\tau_s$ and preserving the energy resolution. However, pulse shaping results in slower pulse onset and consequently reduced peak separation and reduced counting rates \cite{amptek2017sddres}. Using function $f(t)$ (Eq.~\ref{eq2}) for pulse fitting allows keeping both the high energy resolution and high counting rate while eliminating the effect of variations in peaking times.

\subsection{Virtual Multichannel Spectra from Single SDD\label{sec3d}}
\begin{figure*}[!t]
\centering
\includegraphics[width=\textwidth]{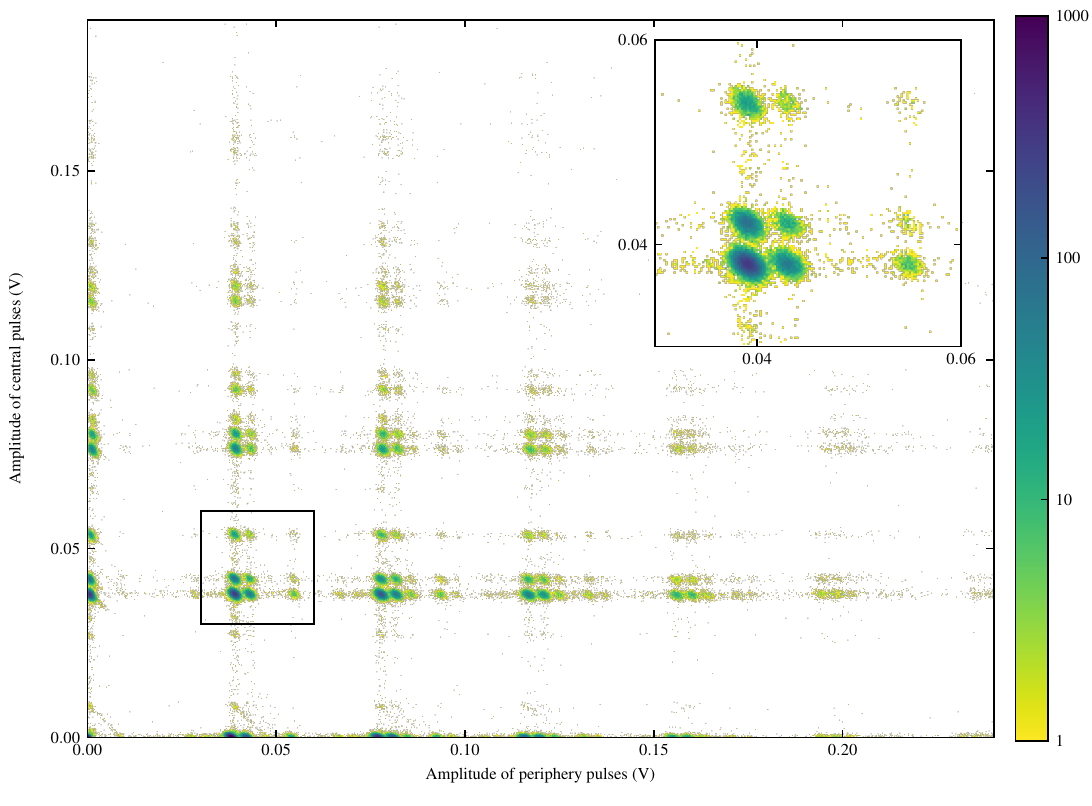}
% where an .eps filename suffix will be assumed under latex, 
% and a .pdf suffix will be assumed for pdflatex; or what has been declared
% via \DeclareGraphicsExtensions.
\caption{Each waveform fit yields two amplitudes, $A^c$ and $A^p$, corresponding to photons entering the SDD through the central and peripheral apertures; here we show an image (on logarithmic scale) of the bidimensional histogram of these values, $A^c$ along y axis, $A^p$ along x axis); different numbers of photons (\numrange{0}{5}~photons along x, \numrange{0}{3}~photons along y) with different energies (sampled from Fe~K\textalpha{ }at \SI{6.40}{\kilo\electronvolt}, Fe~K\textbeta{ }at \SI{7.06}{\kilo\electronvolt}, and FEL beam at $\approx$~\SI{9}{\kilo\electronvolt}) enter through the two apertures yielding a complex histogram; however, the pile-up results in clearly separable ellipsoids from which the number of photons of each individual energy entering through each of the two apertures can be unambiguously reconstructed; the ellipsoids are elongated along an axis at $3\pi/4$, meaning that the sum of the two amplitudes is typically more accurate than the individual amplitudes or their difference; the inset shows a close-up of the area corresponding to one photon entering through each of the two apertures at the \num{3} different energies (yielding 9 ellipsoids).}
\label{fig6}
\end{figure*}

Signal amplitudes $A^c$, $A^p$ resulting from fitting measurements with Eq.~\ref{eq7} are shown in Fig.~\ref{fig6} with a bi-dimensional histogram of the two amplitudes ($A^p$ along the x axis, $A^c$ along the y axis, logarithmic z axis). Multiple clusters of different combinations of energies can be observed.

Fig.~\ref{fig6} inset shows the area corresponding to \num{1}~photon entering through each of the two apertures (in total, \num{2}~photons). Three lines can be identified along each axis, with energies  Fe~K\textalpha{ }(\SI{6.40}{\kilo\electronvolt}), Fe~K\textbeta{ }(\SI{7.06}{\kilo\electronvolt}), and the nominal (\SI{9}{\kilo\electronvolt}) FEL line. Pile-up of \num{2}~photons sampled from \num{3}~lines each yields $3^2=9$~ellipsoids.

The pattern corresponding to Fig.~\ref{fig6}~inset repeats, with increasing complexity and (for this photon rate) decreasing intensity, for higher numbers of photons along each of the axes in Fig.~\ref{fig6}. However, the pile-up results in clearly separable ellipsoids from each combination of (1)~number of photons of each individual energy entering through (2)~each of the 2 apertures.

The ellipsoids are more elongated along an axis at $3\pi/4$ (rotated counterclockwise from the x axis), meaning that the sum of the two amplitudes is typically more accurate than the individual amplitudes or their difference. This is to be expected, as the sum of the amplitudes for signals entering through both apertures can be evaluated using many more samples compared to the difference of the amplitudes (limited to the relatively small set of samples between $t_0^c$ and $t_0^p$, see for example Fig.~\ref{fig3}).

\subsection{Gain and Offset Calibration\label{sec3e}}
To calibrate the energy of the SDD signals, we fitted the positions of the peaks corresponding to \num{0}, Fe~K\textalpha{ }and Fe~K\textbeta{ }(three highest peaks) in the signal amplitude histograms, shown in Fig.~\ref{fig5}~(a), for both the central and peripheral areas of the SDD. The resulting gain and offset values are summarized in Table~\ref{table1}. The central gain and peripheral gains are similar, with a ratio of $\num{0.994} \pm \num{0.001}$.
\begin{table}[!t]
  \renewcommand{\arraystretch}{1.3}
  \caption{Gain and Offset Calibration}
  \label{table1}
  \centering
  \begin{tabular}{{c}{c}{c}}
    \hline
     Aperture & Gain & Offset \\
      & (\SI{E-3}{\volt\per\kilo\electronvolt})& (\SI{E-3}{\volt}) \\
    \hline
    Central & \num{5.948} $\pm$ \num{0.002} & \num{-0.079} $\pm$ \num{0.006} \\
    Peripheral & \num{5.987} $\pm$ \num{0.002} & \num{0.180} $\pm$ \num{0.011} \\
    \hline
  \end{tabular}
\end{table}

\section{Photon Pile-up and Decomposition}
The signal amplitudes $A^c$, $A^p$ resulting from fitting with Eq.~\ref{eq7} were scaled with the corresponding gains and offsets in Table~\ref{table1} and histogrammed to calculate the corresponding spectra.

\subsection{Single Photon Spectrum\label{sec4a}}

\begin{figure}[!t]
\centering
\includegraphics[width=\columnwidth]{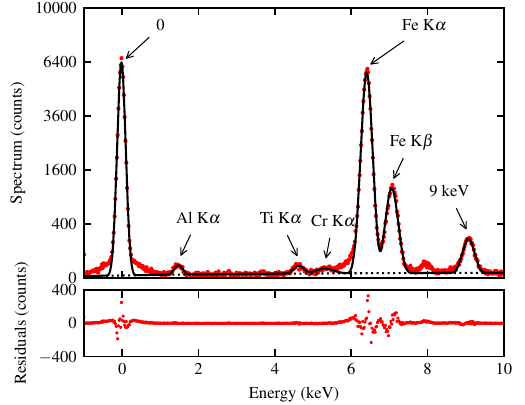}
% where an .eps filename suffix will be assumed under latex, 
% and a .pdf suffix will be assumed for pdflatex; or what has been declared
% via \DeclareGraphicsExtensions.
\caption{Red dots show, on a square root y scale, the spectrum of the photons entering through both apertures up to \SI{10}{\kilo\electronvolt}; this plot reveals three more discrete energies (between \num{0} and Fe~K\textalpha) in addition to the three expected ones; black line depicts the least squares fit using 7 Gaussian peaks and a second degree polynomial background (dotted line); this allows us to identify the three small peaks (as Al~K\textalpha, Ti~K\textalpha, and Cr~K\textalpha), and determine the beam energy (\SI{9.06}{\kilo\electronvolt}); the small peak at $\approx$~\SI{8}{\kilo\electronvolt} is not a new peak, but the pile-up of the Al~K\textalpha{ }and Fe~K\textalpha{ }peaks; while also K\textbeta{ }peaks of Al, Ti and Cr must be present, their contribution is small and can be neglected.}
\label{fig7}
\end{figure}

In Fig.~\ref{fig7} we show a spectrum of (mostly) single photons (up to \SI{10}{\kilo\electronvolt}) entering through either aperture (red dots). The plot is shown on a square root scale to facilitate the simultaneous visualisation of both intense and weak lines. Along the 3 expected lines, there are a few other lines present.

We fitted this spectrum with a simple model of 7 independent Gaussian peaks (6 fundamental lines plus 1 zero peak) and a third degree polynomial background. This allows us to measure the positions of the peaks, their widths and their areas. Table~\ref{table2} lists the fitting parameters and their errors.
\begin{table}[!t]
  \renewcommand{\arraystretch}{1.3}
  \caption{Fitting Parameters for 1 Photon Spectrum}
  \label{table2}
  \centering
  \begin{tabular}{{c}{c}{c}{c}{c}{c}}
    \hline
     Peak & Energy & Peak area & Error & $\sigma$ & $\sigma_{error}$ \\
      & (\si{\kilo\electronvolt}) & (counts) & (counts) & (\si{\electronvolt}) & (\si{\electronvolt}) \\
     \hline
    0 & \num{0.00} & \num{62452} & \num{499} & \num{ 79.78} & \num{  0.48} \\
    Al K$\alpha$ & \num{1.49} & \num{  225} & \num{ 41} & \num{ 89} & \num{ 16} \\
    Ti K$\alpha$ & \num{4.63} & \num{  239} & \num{ 47} & \num{109} & \num{ 22} \\
    Cr K$\alpha$ & \num{5.31} & \num{  239} & \num{ 55} & \num{200} & \num{ 55} \\
    Fe K$\alpha$ & \num{6.40} & \num{79738} & \num{565} & \num{112.59} & \num{  0.63} \\
    Fe K$\beta$ & \num{7.06} & \num{16511} & \num{265} & \num{122.6} & \num{  1.7} \\
    9 keV & \num{9.06} & \num{ 3022} & \num{121} & \num{119.7} & \num{  4.3} \\
    \hline
    \multicolumn{6}{l}{$\chi_{\nu}^2 = \num{  73.4}$ (reduced chi squared)}\\
  \end{tabular}
\end{table}

Using the peak positions, we identified the six fundamental lines as: Al~K\textalpha, Ti~K\textalpha, Cr~K\textalpha, Fe~K\textalpha, Fe~K\textbeta, and the nominal \SI{9}{\kilo\electronvolt} line from the FEL beam, and labeled the peaks accordingly. We also determined the beam energy (\SI{9.06}{\kilo\electronvolt}).

The small peak at $\approx$~\SI{8}{\kilo\electronvolt} does not represent a new fundamental line; it is the pile-up of the Al~K\textalpha{ }and Fe~K\textalpha{ }\num{1}~photon peaks. While also K\textbeta{ }peaks of Al, Ti and Cr must be present, their yield is much lower than the corresponding K\textalpha{ }yield; multiplied with their low K\textalpha{ }intensity, their contribution can be neglected.

The peak areas in Table~\ref{table2} provide a first indication of the intensities of the different peaks; however, using peak areas as a proxies for  average detection rates yields accurate rate estimations only when each of the $\lambda_i$ rates is much smaller than \num{1}~photon/event (subsection~\ref{sec2f}) and ignoring the FEL distribution of beam intensities (subsection~\ref{sec2g}). The peak width estimates have large estimated errors in the peaks with low statistics, and the reduced chi-squared statistic is relatively large ($\chi_{\nu}^2=73.4$), due to the ``heavy tails''.

This simple, typical analysis illustrates the limitations of using only \num{1}~photon peaks. In next subsection (\ref{sec4b}) we will address these limitations with an appropriate pile-up fitting model.

\subsection{Photon Pile-up Spectrum and Decomposition\label{sec4b}}

\begin{figure*}[!t]
\centering
\includegraphics[width=\textwidth]{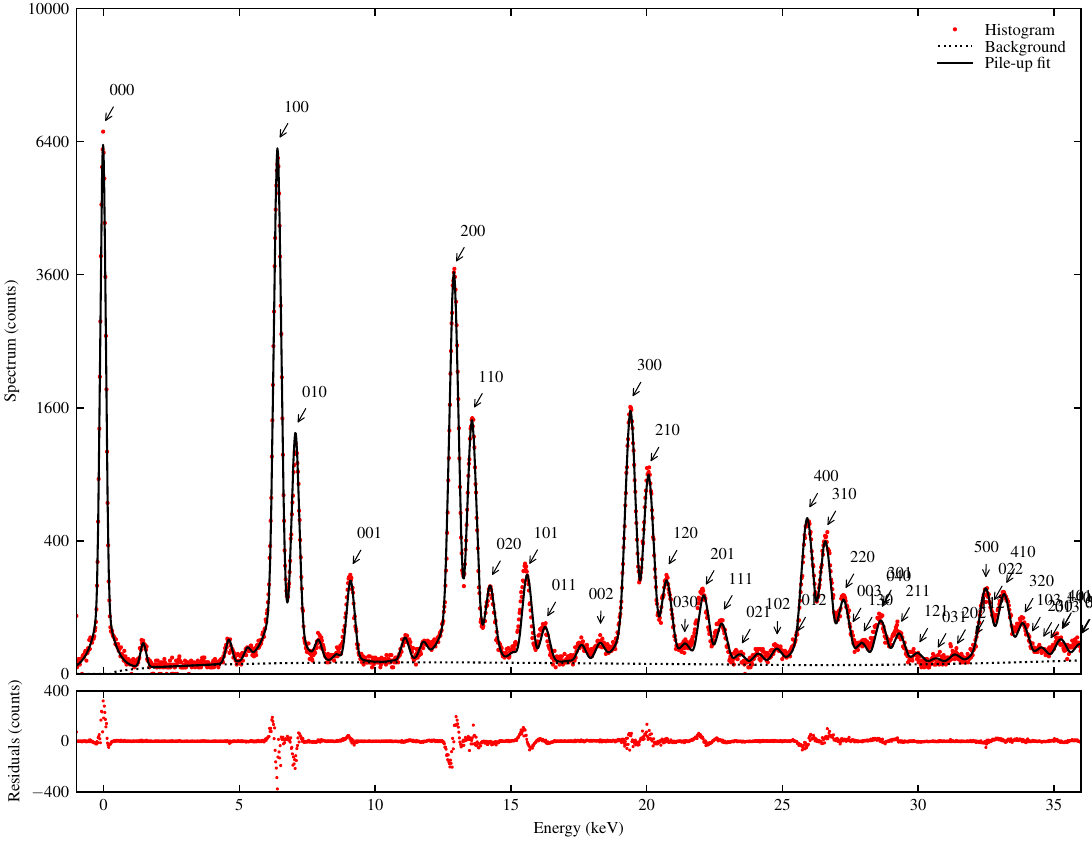}
% where an .eps filename suffix will be assumed under latex, 
% and a .pdf suffix will be assumed for pdflatex; or what has been declared
% via \DeclareGraphicsExtensions.
\caption{Red dots show (on a square root y scale) the spectrum of photons entering through both apertures; black line depicts the least squares fit using the photon pile-up model (Eq.~\ref{eq26} including Eq.~\ref{eq13},~\ref{eq18},~\ref{eq27}) with the theoretical peak positions of the 7 fundamental lines (Al~K\textalpha, Ti~K\textalpha, Cr~K\textalpha, Fe~K\textalpha, Fe~K\textbeta, and \SI{9.06}{\kilo\electronvolt}); fitting yields their individual photon rates $\lambda_i$ (effectively decomposing the photon  pile-up), noise parameters $\sigma_0$ and $\sigma_1$ (Eq.~\ref{eq15}), shape parameter $\alpha$ of the gamma distribution of FEL pulse intensities, ``heavy tail'' parameters $\beta$ and $\eta$, and a third degree polynomial background (indicated by the dotted  black line); the photon pile-up model is remarkably successful in fitting the plethora of peaks, using only the position of fundamental peaks and mathematical modeling; the FEL beam does not have a constant intensity, resulting in a stretching of the Poisson distribution of number of counts (parametrized by Eq.~\ref{eq13}); for the 3 most intense lines (Fe~K\textalpha, Fe~K\textbeta, and \SI{9.06}{keV}) we indexed the peaks with corresponding 3 digit labels (representing the number of photons from each of the 3 lines), showing that the photon pile-up model (subsection~\ref{sec2i}) accurately describes the measured spectrum ($\chi_{\nu}^2=3.71$); overlapping labels indicate overlapping peaks.}
\label{fig8}
\end{figure*}

Fig.~\ref{fig8} depicts the entire spectrum of photons entering through either aperture (red dots), up to \SI{37}{\kilo\electronvolt}; to facilitate inspection of both intense and weak lines, the plot uses a square root y axis. The \num{6} fundamental lines (single photons) are clearly visible. Pile-up of multiple photons (up to \num{5} shown) results in increasing complexity of the detected spectrum and decreasing histogram height (in this particular case).

The black line depicts the least squares fit using a fitting model with theoretical peak positions of the 7 fundamental lines (zero peak, Al~K\textalpha, Ti~K\textalpha, Cr~K\textalpha, Fe~K\textalpha, Fe~K\textbeta, and \SI{9.06}{keV}), using Eq.~\ref{eq26} with the appropriate changes (substituting Eq.~\ref{eq14} and \ref{eq18} for Eq.~\ref{eq8} and \ref{eq16}, respectively, to account for the distribution of the FEL beam intensity and the observed ``heavy tails'' of peaks).

The fit accurately reproduces most features in the data set, resulting in a small reduced chi-squared statistic $\chi_{\nu}^2=3.71$ and reproducing all characteristics of the plethora of peaks. For the 3 most intense lines (Fe~K\textalpha, Fe~K\textbeta, and \SI{9.06}{keV}) we indexed the peaks with corresponding 3 digit labels. These peaks dominate the higher energy pile-up and can still be clearly separated.
\begin{table}[!t]
  \renewcommand{\arraystretch}{1.3}
  \caption{Fit Parameters for Photon Pile-up Spectrum}
  \label{table3}
  \centering
  \begin{tabular}{{r}{r}{r}{l}{l}}
    \hline
     Parameter & Value & Error & Unit & Details \\
    \hline
    $\sigma_0$ & \num{ 77.28} & \num{  0.44} & \si{\electronvolt} & electronic noise \\
    $\sigma_1$ & \num{  0.7700} & \num{  0.0089} & \si{\electronvolt} & energy dependent noise \\
    $\alpha$ & \num{ 17.9} & \num{  1.3} &  & $\Gamma_{PDF}$ shape \\
    $\beta$ & \num{  0.798} & \num{  0.019} &  & $G^*$ heavy tail shape \\
    $\eta$ & \num{  0.0914} & \num{  0.0052} &  & $G^*$ heavy tail fraction \\
    $\lambda_{Al K\alpha}$ & \num{  0.0035} & \num{  0.0003} & photons/event & detection rate \\
    $\lambda_{Ti K\alpha}$ & \num{  0.0048} & \num{  0.0003} & photons/event & detection rate \\
    $\lambda_{Cr K\alpha}$ & \num{  0.0022} & \num{  0.0003} & photons/event & detection rate \\
    $\lambda_{Fe K\alpha}$ & \num{  1.4322} & \num{  0.0041} & photons/event & detection rate \\
    $\lambda_{Fe K\beta}$ & \num{  0.2869} & \num{  0.0018} & photons/event & detection rate \\
    $\lambda_{9 keV}$ & \num{  0.0450} & \num{  0.0008} & photons/event & detection rate \\
    \hline
\multicolumn{5}{l}{$\chi_{\nu}^2 = \num{   3.71}$ (reduced chi squared)} \\
  \end{tabular}
\end{table}

\subsubsection{Average photon rate}
Fitting the pile-up spectrum with the pile-up model yields estimates of photon rates in individual lines (Table~\ref{table3}). For peaks with good statistics, the relative error is small (\SI{0.23}{\percent} for Fe~K\textalpha{ }and \SI{0.63}{\percent} for Fe~K\textbeta), increasing to \SI{14}{\percent} for the peak with lowest statistics, Cr~K\textalpha.

For the entire spectrum, the most likely photon pile-up decomposition results in a number of photons $N\lambda_i$ from each line $i$, where $N$ is the number of events. The average rate for all photons $\lambda=\sum \lambda_i = \num{1.7746}\pm{0.0046}$~photons/event, with a majority of Fe~K\textalpha{ }photons ($\num{1.4322}\pm\num{0.0041}$~photons/event).

\subsubsection{Noise\label{sec4b2}}
For Mn~K\textalpha{ }monochromatic radiation (i.e., \SI{5.89}{\kilo\electronvolt}), we obtain a line width of $2.355 \sqrt{\sigma_0^2+\sigma_1\cdot\SI{5.89}{\kilo\electronvolt}}=\SI{241}{\electronvolt}$~FWHM.

On average, we observe a fraction $\eta=\SI{9.1}{\percent}$ of photons in the ``heavy tails'', with a distribution shape parameter $\beta=0.798$ (details in subsection~\ref{sec2h}).

The ``heavy tails'' and a significant fraction of noise are most likely a consequence of severe undersampling of the analog prefilter output: the step times (or rise times) we observed in the absence of pulse shaping  ($\tau_s^c=\SI{21.0}{\nano\second}$ and $\tau_s^p=\SI{43.3}{\nano\second}$) were much shorter than the sampling period (\SI{100}{\nano\second}, corresponding to \SI{10}{\mega\hertz} ADC rate of the PX5, see Appendix for details). While fitting evaluates pulse height accurately despite undersampling, the optimal estimator for detected energy is signal amplitude (subsection~\ref{sec3c}), thus imprecision in determining step times $\tau_s$ results in loss of energy resolution (as described in subsection~\ref{sec2d} by Eq.~\ref{eq4}).

We expect the energy resolution of the pulse fitting method presented here to approach the theoretical $\SI{128}{\electronvolt}$~FWHM with improved sampling ($\ge$~\SI{100}{\mega\hertz}) while maintaining the pulse separation (\SI{720}{\nano\second}, similar to a counting rate of \SI{1.4}{\mega\count\per\second} and no rejection).

Current pulse processors can be tuned to yield a high counting rate, or a high energy resolution, or a compromise between the two, by choosing an appropriate peaking time. The Amptek XR-100SDD and PX5 digital pulse processor can yield an energy resolution of \SI{240}{\electronvolt} at Mn~K\textalpha{ }with a peaking time of $\approx$~\SI{0.1}{\micro\second} and acquisition rate of \SI{80}{\mega\hertz} \cite{amptek2017sddres}, resulting in a maximum output count rate of $\approx$~\SI{1}{\mega\count\per\second} \cite{amptek2017sddres} with $\approx$~\SI{60}{\percent} rejection rate \cite{amptek2017sddres}; we demonstrated improved results with function fitting of severely undersampled waveforms.

The Amptek XR-100SDD and PX5 digital pulse processor can also yield an energy resolution of \SI{130}{\electronvolt} at Mn~K\textalpha{ }with a peaking time of \SI{10}{\micro\second} and acquisition rate of \SI{80}{\mega\hertz} \cite{amptek2017sddres}, resulting in a maximum output count rate of \SI{20}{\kilo\count\per\second} \cite{amptek2017sddres} with $\approx$~\SI{60}{\percent} rejection rate \cite{amptek2017sddres}. With optimal sampling ($\ge$~\SI{100}{\mega\hertz\per\second}), we expect $\approx$~\num{2}~orders of magnitude higher counting rates than current pulse processing approaches while achieving close to the theoretical limit in energy resolution.

\subsubsection{FEL statistics}
Using the appropriate model (Eq.~\ref{eq13}) which takes into account the gamma distribution of FEL pulses (subsection~\ref{sec2g}), we obtain an intensity distribution parameter $\alpha=17.9$. Note that this parameter is specific to the current experiment and depends on beam characteristics, optics, etc. For reference, a constant intensity beam has a shape $\alpha \to \infty$. %The small mismatch between the histogram and fit in the 000 peak is due to occasionally dropped pulses of the FEL (which are not described by the gamma distribution).

\subsection{Single Event Bayesian Decomposition of Photon Pile-up}
The maximum likelihood decomposition and error estimates depend strongly on the detected spectrum; in this subsection we use the spectrum characteristics obtained in previous subsection (\ref{sec4b}) and detailed in Table~\ref{table3}; we ignored in these estimates the ``heavy tails'' to estimate error rates that are representative for most SDD setups (including the ``heavy tails'' increases the error rates by a factor of \num{2.4}).

\subsubsection{Pile-up peak error\label{sec4c1}}
\begin{figure}[!t]
\centering
\includegraphics[width=\columnwidth]{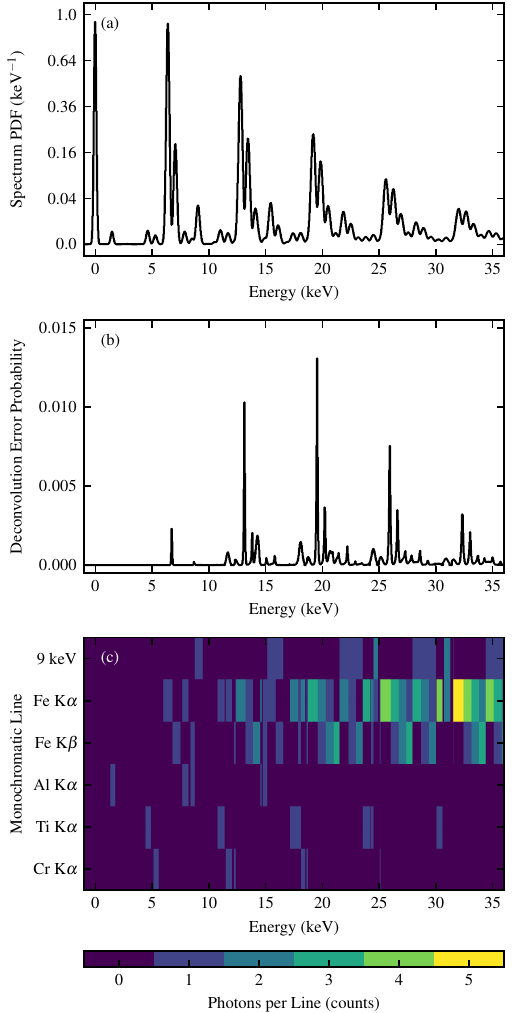}
% where an .eps filename suffix will be assumed under latex, 
% and a .pdf suffix will be assumed for pdflatex; or what has been declared
% via \DeclareGraphicsExtensions.
\caption{Pile-up deconvolution of single events: (a)~probability density function (PDF) of the incoming spectrum, shown on a square root scale; (b)~probability of error in choosing the most likely pile-up peak for each event energy; its integral represents the total error rate, which for this spectrum is $\SI{1.11}{\percent}$; (c)~map showing the most likely combinations of numbers of photons in each line for each detected energy.}
\label{fig9}
\end{figure}

Fig.~\ref{fig9} shows a single event decomposition approach. Using the method described in subsection~\ref{sec2h}, we calculated the probability $S(E)$ of each energy $E$, depicted in Fig.~\ref{fig9}~(a), and with the method described in subsection~\ref{sec2k2} the error probability $Err_{peak}(E)$, shown in Fig.~\ref{fig9}~(b). The error rate of peak identification is \SI{1.11}{\percent}.

\subsubsection{Maximum likelihood decomposition\label{sec4c2}}
For each energy $E$ we estimated the maximum likelihood decomposition in numbers of photons from each line, described in subsection~\ref{sec2k1} and indicated in Fig.~\ref{fig9}~(c). Note that both the maximum likelihood decomposition and error estimate depend strongly on the spectrum characteristics (summarized in Table~\ref{table3}).

\subsubsection{Photon error\label{sec4c3}}
To estimate the error rate in decomposition of photon pile-up, we simulated \num{E6}~events with characteristics described by Table~\ref{table3}, using the method described in subsection~\ref{sec2k3}, generating numbers of photons, calculating their pile-up, adding the appropriate noise and then using the maximum likelihood decomposition in subsection~\ref{sec2k1}.

We collected the results, for each event, in a confusion matrix, shown in Table~\ref{table4}; rows correspond to simulated photons, while columns correspond to the maximum likelihood decomposition. Cell $(i,j)$ represents how many photons from line $i$ were allocated to line $j$. Successfully decomposed photons are added to the diagonal. The error rate of photon allocation is \SI{1.05}{\percent}. Assuming the same spectrum and a detector operating at the Fano limit, we obtain an error rate of photon allocation of \SI{0.64}{\percent}.
\begin{table}[!t]
  \renewcommand{\arraystretch}{1.3}
  \caption{Confusion matrix of single event decomposition using \num{E+6} simulated events}
  \label{table4}
  \centering
  \begin{tabular}{{c}{c}{c}{c}{c}{c}{c}}
     & Al K$\alpha$ & Ti K$\alpha$ & Cr K$\alpha$ & Fe K$\alpha$ & Fe K$\beta$ & 9 keV \\
    \hline
    Al K$\alpha$ & \num{1913} & \num{17} & \num{0} & \num{109} & \num{1258} & \num{173}\\
    Ti K$\alpha$ & \num{0} & \num{3564} & \num{231} & \num{423} & \num{119} & \num{297}\\
    Cr K$\alpha$ & \num{2} & \num{84} & \num{1216} & \num{388} & \num{154} & \num{184}\\
    Fe K$\alpha$ & \num{0} & \num{78} & \num{27} & \num{1372827} & \num{6131} & \num{638}\\
    Fe K$\beta$ & \num{18} & \num{32} & \num{231} & \num{5420} & \num{271348} & \num{683}\\
    9 keV & \num{20} & \num{122} & \num{90} & \num{620} & \num{381} & \num{41980}\\
    \hline
    Sum  & \num{1953}  & \num{3897}  & \num{1795}  & \num{1379786}  & \num{279390}  & \num{43955}  \\
    \hline
  \end{tabular}
\end{table}

\subsubsection{Using multiple apertures\label{sec4c4}}
Splitting the pile-up signal in two independent signals using the apertures reduces the photon error rate by avoiding pile-up of large numbers of photons, where the density of pile-up peaks becomes higher. Using two apertures we obtained a reduction in error rate of photon allocation to \SI{0.55}{\percent} (and \SI{0.31}{\percent} with a detector at the Fano limit).

Splitting the pile-up signal in two also reduces the number of pile-up peaks to consider, increasing computation speed; e.g., for the spectrum discussed here, using $M=6$~photons guarantees that less than \num{E-4} of the events have more than $M$ counts in one line. Splitting the pile-up signal in two results in a maximum number of photons of $M=4$. With $L=6$ lines, this is a reduction from \num{924} to \num{210} pile-up peaks while doubling the number of calculations, thus speeding up fitting by a factor \num{2.2}.

\section{ePixS Pixel Detectors: SDD Successors}

ePixS \cite{hasi2015high,dragone2015epixs} is a 2D, charge integrating, hybrid pixel detector with an array of \num{10x10} \SI{500x500}{\micro\metre} pixels, where each pixel has a spectroscopic performance (noise $\sigma$~of \SI{8}{\electron} or $\approx$~\SI{30}{\electronvolt}, and measured line width of $\approx$~\SI{215}{\electronvolt} FWHM at Mn~K\textalpha) approaching that of SDDs. The ePixS camera is built on the ePix platform \cite{dragone2013epix}, providing a large number of spectroscopic imaging pixels in a compact, robust and affordable camera package \cite{nishimura2015design}.

\begin{figure}[!t]
\centering
\includegraphics[width=\columnwidth]{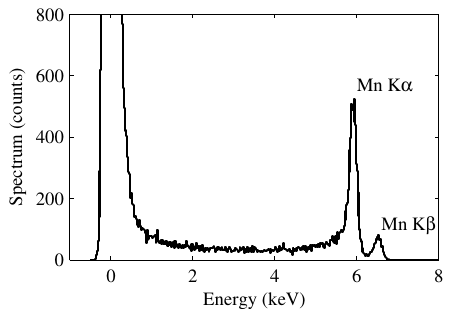}
%(b)\includegraphics[width=\columnwidth]{fig10b.eps}
% where an .eps filename suffix will be assumed under latex, 
% and a .pdf suffix will be assumed for pdflatex; or what has been declared
% via \DeclareGraphicsExtensions.
\caption{The ePixS spectroscopic, hybrid pixel detector (reproduced with permission from \cite{nishimura2015design}): ePixS spectrum obtained over all pixels using an \(^{55}\text{Fe}\) source, generating characteristic Mn~K\textalpha{ }(\SI{5.9}{\kilo\electronvolt}) and Mn~K\textbeta{ }(\SI{6.5}{\kilo\electronvolt}) photons, showing the low noise operation with a resulting line width of \SI{215}{\electronvolt} at Mn~K\textalpha.}
\label{fig10}
\end{figure}

Fig.~\ref{fig10} shows an \(^{55}\text{Fe}\) source spectrum (Mn~K\textalpha~at \SI{5.90}{\kilo\electronvolt} and Mn~K\textbeta~at \SI{6.49}{\kilo\electronvolt} lines) measured with an ePixS detector (reproduced with permission from \cite{nishimura2015design}). The spectrum is collected in all pixels individually and shown in a single plot here. The ePixS energy resolution is similar to the energy resolution we observed for the Amptek SDD in subsection~\ref{sec4b}.

ePixS is a charge integrating pixel detector, providing only the total signal collected by each pixel in each FEL pulse. As a direct result, the pulse processing described in subsections~\ref{sec2c}~to~\ref{sec2e} is not applicable to ePixS.

However, with the same area as the single channel SDD presented here (\SI{25}{\milli\metre\squared}) and \num{100} independent pixels (i.e., position sensitive independent channels) with similar energy resolution, an ePixS detector can measure much higher photon rates with limited pile-up and similar spectroscopic performance, while being intrinsically position-sensitive.

For the remaining pile-up, fitting and decomposition described in subsections~\ref{sec2f}~to~\ref{sec2k} are applicable, yielding similar results to those presented here for SDDs: \num{100}~pixels allow acquisitions of global rates of $\lambda \approx 178$~photons/pulse while allowing decomposition into single photons with $\approx\SI{99}{\percent}$ accuracy (error rate of \SI{1.05}{\percent}) and yielding their position.

\section{Conclusion}

In standard spectroscopy applications there is often a complex spectrum with a multitude of peaks which must be resolved and matched to specific elements.

In contrast, at x-ray FELs there are often only a few photon energies involved in an experiment, with significant photon pile-up. The discrete spectrum significantly reduces the complexity of pile-up decomposition. SDDs can bring significant contributions at FELs in (1)~photon counting, (2)~spectrum measurements, and (3)~background measurements and set-up optimization.

We present a pulse processing method based on fitting individual, overlapping pulses with pulse functions (a combination of gradual step function and exponential decay) which allows extracting (in the time domain) precise amplitude and timing information in the presence of pulse pile-up and clipping.

We demonstrate that, in the absence of pulse shaping, the pulse height or pulse area are not optimal estimators for the detected photon energy (due to variation in peaking times). Instead, the optimal estimator is signal amplitude, obtained (together with the rise time) by fitting each pulse waveform with the pulse function. This removes the need for pulse shaping, resulting in fast pulse onset and improved pulse separation, allowing fitting of multiple pulse pile-up.

We show that partially overlapping peaks can be separated, down to differences in time-of-arrival of $\approx$~\SI{700}{\nano\second} (corresponding to $\approx$~\SI{1.4}{\mega\count\per\second} and no rejection), while yielding pulse time-of-arrival with precision of \SI{10}{\nano\second}.

Typical pulse processing involves a compromise between counting rate (with short rise times) and energy resolution (long rise times through pulse shaping) by choosing an appropriate peaking time.

Both high energy resolution and high counting rates can be achieved by fitting the waveform in the absence of pulse shaping. We demonstrated improved results compared to current pulse processing, despite severely undersampled waveforms. With improved sampling, we expect a high energy resolution ($\approx$~\SI{130}{\kilo\electronvolt}) while achieving $\approx$~\num{2}~orders of magnitude higher counting rates than current pulse processing methods.

While we demonstrate the usage of this pulse processing method with an SDD and pulsed FEL source, it can be extended to any detector with rapid response (e.g., SDD, transition edge sensors) and x-ray source (e.g., FEL, synchrotron, x-ray tube). In SDDs at pulsed sources, the timing can be used to recover the interaction radii.

At pulsed sources, or at short time intervals between two photons, photon pile-up occurs and the photon rates are not accurately described by the standard proxy, i.e., area of one photon peaks.

Instead, the photon pile-up of a discrete detected spectrum can be accurately described and fitted with the photon pile-up model presented here, yielding precise estimates of the photon rates in individual lines and effectively decomposing the spectrum. We also extended the model to include the stretching of the Poissonian statistics of pile-up introduced by sources with variable intensity (e.g., FELs).

We present a Bayesian decomposition approach which allows accurate decomposition of individual photon energies in single pile-up events, and estimating its error rate in the order of \SI{1}{\percent} (\SI{99}{\percent} accuracy) for the spectrum discussed in this paper (average rate $\lambda=1.78$~photons/event, with pile-up of up to \num{6}~photons from \num{6} monochromatic lines).

The photon pile-up and Bayesian decomposition presented here are useful not only for SDDs but for any applications of pile-up decomposition (e.g., spectroscopy with transition edge sensors \cite{irwin2005transition}, low-noise spectroscopic imaging with integrating pixel detectors \cite{blaj2016xray}), or even usual pulse processing with pile-up rejection (two photons with small differences in time-of-arrival will not be distinguished by the pile-up rejection, resulting in pile-up).

The usefulness of silicon drift detectors will continue into the x-ray FEL era of science. Their successors, the ePixS spectroscopic, hybrid pixel detectors already offer hundreds of pixels with similar performance in a compact, robust and affordable package, particularly useful in x-ray FELs \cite{blaj2015future}.

\section*{Glossary}
%\begin{table}[!t]
%\renewcommand{\arraystretch}{1.3}
% if using array.sty, it might be a good idea to tweak the value of
% \extrarowheight as needed to properly center the text within the cells
%\caption{An Example of a Table}
%\label{tabglossary}
%\centering
%% Some packages, such as MDW tools, offer better commands for making tables
%% than the plain LaTeX2e tabular which is used here.
\begin{tabular}{r l}
ADC: & Analog-to-digital converter\\
CSPAD: & Cornell-SLAC pixel array detector\\
CXI: & Coherent X-ray Imaging instrument at LCLS\\
DAC: & Digital-to-analog converter\\
ePix: & SLAC hybrid pixel detector platform\\
ePixS: & Spectroscopic pixel detector in the ePix family\\
FEL: & Free-electron laser\\
FWHM: & Full width at half maximum ($\approx$~2.355~$\sigma$)\\
LCLS: & Linac Coherent Light Source at SLAC\\
PDF: & Probability density function\\
r.m.s.: & Root mean square\\
SDD: & Silicon drift detector\\
SLAC: & SLAC National Accelerator Laboratory\\
XPP: & X-ray Pump-Probe instrument at LCLS\\
%\hline
%One & Two\\
%\hline
%Three & Four\\
%\hline
\end{tabular}
%\end{table}

% or


\appendix[Waveform processing\label{appendix}]

\begin{figure}[!t]
\centering
\includegraphics[width=\columnwidth]{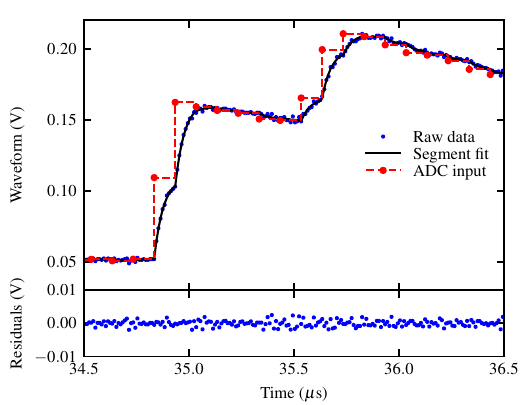}
% where an .eps filename suffix will be assumed under latex, 
% and a .pdf suffix will be assumed for pdflatex; or what has been declared
% via \DeclareGraphicsExtensions.
\caption{Blue dots indicate a typical raw waveform of the Amptek PX5 analog prefilter, acquired at \SI{100}{\mega\hertz}; the Amptek PX5 DAC only updates at \SI{10}{\mega\hertz} \cite{amptek2016px5}, resulting in a series of \num{10}~sample segments, each with an exponential decay towards its DAC target value; the black lines indicate the least squares fit for each segment, and the dashed red line with large red dots indicate the reconstruction of the DAC output, representing the analog prefilter output of the Amptek PX5; arrival times of $\approx\SI{35}{\micro\second}$ are calculated from arrival time $t_0=0$ of the FEL trigger signal.}
\label{fig11}
\end{figure}

The AUX1 analog signal of PX5 does not provide the analog prefilter output signal directly \cite{amptek2016px5}. Instead, it is digitized by the ADC at \SI{80}{\mega\hertz} and subsequently synthesized by the DAC at \SI{10}{\mega\hertz} \cite{amptek2016px5}. This is clearly visible in the raw data, shown with small blue dots in Fig.~\ref{fig11}; each DAC change results in a transient signal with an exponential decay from the current output towards the target output. The time constant of the DAC exponential low pas filter is $\tau=\SI{43.5}{\nano\second}$ in our setup.

To recover the initial signal from the transient exponential decay segments, we fitted each \SI{100}{\nano\second} (\num{10} samples) segment of raw data between 2 DAC updates with an exponential decay:
\begin{align} \label{eq31}
    y_{raw}=y_{ADC}+A \e^{-\frac{t-t_o}{\tau}}
\end{align}
where $t_o$ is the time offset of the first sample in the current segment, $y_{ADC}$ the new target output. The PX5 DAC and Acquiris were not synchronized and their time offset changed in each waveform. However, the offset can be recovered from the data by using $t_o$ as a global waveform fitting parameter and minimizing $\chi^2$.

The resulting fit is indicated with a black line for one waveform in Fig.~\ref{fig11}. The dashed red line and large red dots indicate the reconstructed time and amplitude of the DAC output, correcting for the system low-pass filtering. Correcting distortions introduced by the ADC, DAC and exponential low-pass filtering results in significantly improved results.

Arrival times are calculated from time $t=0$ corresponding to the arrival of the trigger signal that precedes each FEL pulse, with $t \approx \SI{35}{\micro\second}$ determined by the experimental layout and cable lengths.

%\appendices
%Some text for the appendix.

\section*{Acknowledgment}

Use of the Linac Coherent Light Source (LCLS), SLAC National Accelerator Laboratory, is supported by the U.S. Department of Energy, Office of Science, Office of Basic Energy Sciences under Contract No. DE-AC02-76SF00515.

We applied the SDC approach for the sequence of authors \cite{tscharntke2007author}. Statement of authorship: conception, C.~J.~Kenney and G.~Blaj; analytical methods, G.~Blaj; analysis, G.~Blaj; LCLS beamtime principal investigator, G.~Carini; ePixS ASIC design, A.~Dragone; design and acquisition of data, all authors; drafting the manuscript, G.~Blaj; revising the manuscript: G.~Blaj and C.~J.~Kenney.

\ifCLASSOPTIONcaptionsoff
  \newpage
\fi

\end{document}